\documentclass[a4paper,11pt]{article}
\usepackage{pos,soul}

\definecolor{MyBlue}{HTML}{0000FE}
\definecolor{MyGrey}{HTML}{999999}
\definecolor{MyOrange}{HTML}{FF9900}
\definecolor{MyDarkRed}{HTML}{CC0033}
\definecolor{MyPurple}{HTML}{A78EC2}
\definecolor{MyDarkGreen}{HTML}{006766}
\definecolor{MyCyan}{HTML}{00D1D1}
\definecolor{MyLightBlue}{HTML}{5785FF}
\definecolor{MyPink}{HTML}{FF00FE}

\newcommand{\RNum}[1]{\uppercase\expandafter{\romannumeral #1\relax}}

\title{Form factors for semileptonic $B_{(s)}\to D^*_{(s)} \ell\nu_\ell$ decays}

\author*{Anastasia Boushmelev}
\author[1]{Matthew Black}
\note{Present address: School of Physics and Astronomy, University of Edinburgh, Edinburgh EH9~3JZ, UK}
\author{Oliver Witzel}

\onbehalf{\newline (RBC/UKQCD collaborations)}
\affiliation{Theoretische Physik 1, Center for Particle Physics Siegen, Universität Siegen, \\57068 Siegen, Germany}

\emailAdd{anastasia.boushmelev@uni-siegen.de}

\abstract{
Semileptonic $B_{(s)}$ decays are of great phenomenological interest
because they allow to extract CKM matrix elements or test lepton flavour
universality. Taking advantage of existing data, we explore extracting
form factors for vector final states using the narrow width
approximation. Based on RBC/UKQCD's set of 2+1 flavour gauge field
ensembles with Shamir domain-wall fermion and Iwasaki gauge field
action, we study semileptonic $B_{(s)}$ decays using domain-wall
fermions for light, strange and charm quarks, whereas bottom quarks are
simulated with the relativistic heavy quark (RHQ) action. Exploratory
results for $B_s \to D_s^* \ell \nu_\ell$ are presented.
}

\FullConference{The 41st International Symposium on Lattice Field Theory (LATTICE2024)\\
 28 July - 3 August 2024\\
Liverpool, UK\\}

\begin{document}
\maketitle

\section{Introduction}

\begin{figure}[b]
\centering
    \begin{picture}(108,30)(-1,40)
    \put(6,43){\includegraphics[width=60mm]{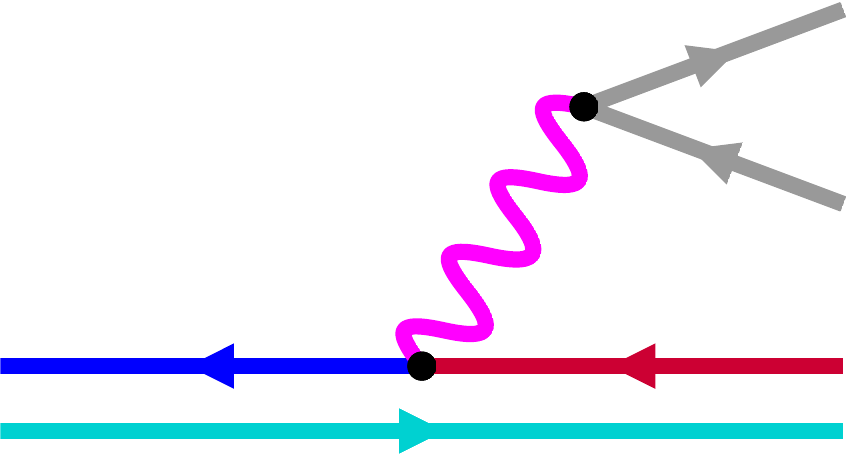}}
    \put(-1,46){\large{$B_{(s)}$}} \put(57,46){\large{$D_{(s)}^*$}}
    \put(28,56){\large{\textcolor{MyPink}{$W$}}} 
    \put(56,67){\large{$\ell$}} \put(56,57){\large{$\nu_\ell$}}
    \put(60,62.7){$\Bigg\}$}
    \put(64,63){$q^2= M_{B_{(s)}}^2 + M_{D_{(s)}^*}^2 - 2 E_{B_{(s)}} E_{D_{(s)}^*}$}
    \put(30,41){\textcolor{MyCyan}{$q$}}
    \put(47,50){\textcolor{MyDarkRed}{$\overline{c}$}}
    \put(18,50){\textcolor{MyBlue}{$\overline{b}$}}
    \end{picture}
    \caption{Feynman diagram depicting tree-level semileptonic decays $B_{(s)}\to D_{(s)}^* \ell \nu_\ell$ for a spectator quark $q=u/d,\,s$ and a lepton-neutrino pair with $\ell=e,\,\mu,\,\tau$. }
    \label{fig:feynmandiag}
\end{figure}
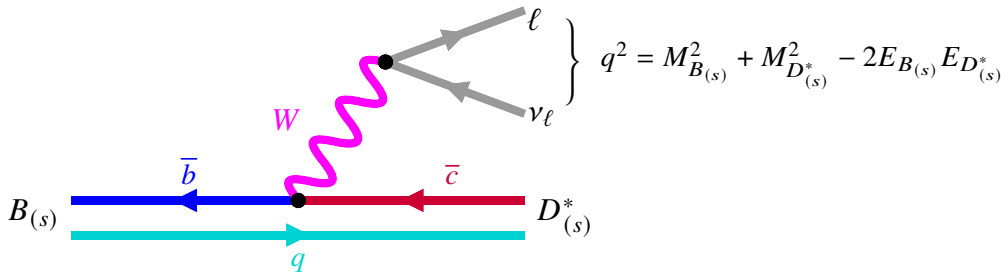

Flavour physics plays an important role in testing and constraining the Standard Model (SM) and guiding the search for new physics beyond the SM. Both experimental measurements and theoretical predictions derived from the SM are required either to extract fundamental parameters of the SM, such as Cabibbo-Kobayashi-Maskawa (CKM) matrix elements, or test properties, like lepton flavour universality (LFU).
  Of special interest in this quest are bottom flavoured quarks because bottom, or $b$ for short, quarks are the heaviest quarks living long enough to form experimentally accessible bound states. \\
 In the following we focus on exclusive semileptonic decays with hadronic vector final states, $B_{(s)}\to D_{(s)}^* \ell \nu_\ell$, which are described by tree-level charged-current processes in the SM. 
 A Feynman diagram of these decays is shown in Fig.~\ref{fig:feynmandiag} where a light or a strange quark acts as ``spectator''.

Although this $b\to c$ transition occurs at tree-level in the SM and is therefore not expected to have large new physics contributions, two observations have attracted the attention of the community for many years:
\begin{enumerate}
    \item ~Tree-level semileptonic decays provide our most precise channels to extract the CKM matrix element $|V_{cb}|$ describing the transition amplitude from a $b$ to $c$ quark. We can either use an inclusive approach summing over all final states containing a charm quark or an exclusive approach explicitly considering a $D$ or $D^*$ meson in the final state. A $2-3 \sigma$ tension between inclusive and exclusive determinations of $|V_{cb}|$ has persisted for several years. Presently the Particle Data Group (PDG) quotes \cite{ParticleDataGroup:2024cfk}
    \begin{align}
    |V_{cb}|^{\text{incl}}=(42.2 \pm 0.5) \times 10^{-3}\qquad\text{and}\qquad |V_{cb}|^{\text{excl}}=(39.8\pm0.6) \times 10^{-3},
    \end{align}
    which results in an $\chi^2$-inflated average with about 4\% uncertainty: 
    $|V_{cb}|=(41.1 \pm 1.2) \times 10^{-3}$.
\item ~Considering different leptonic final states in exclusive $B\to D^{(*)}\ell\nu_\ell$ decays, we can define ratios testing LFU in the SM
    \begin{align}
    \mathcal{R}(D^{(*)})=\dfrac{\mathcal{B}(B\to D^{(*)} \tau \nu_\tau)}{\mathcal{B}(B \to D^{(*)} l\nu_l)}
    \qquad\text{with}\quad l={e,\mu}.
    \end{align}
Here a $2-3\sigma$ tension \cite{HFLAV:RdRds-2024} exists between the theoretical prediction and experimental result.
\end{enumerate}

We will consider semileptonic $B_{(s)}\to D_{(s)}^*\ell\nu_\ell$ decays since spin-1 vector final states are experimentally favoured. Experimental results for these processes have been reported by BaBar \cite{BaBar:2007cke, BaBar:2007ddh}, BELLE \cite{Belle:2018ezy,Belle:2019rba}, BELLE \RNum{2} \cite{Belle-II:2020ewe,Belle-II:2020dyp} and LHCb \cite{LHCb:2015gmp,LHCb:2017smo,LHCb:2022piu}. Using lattice quantum chromodynamics (QCD) we aim to make a nonperturbative determination of the form factors describing these decays. To date Fermilab/MILC \cite{FermilabLattice:2021cdg}, JLQCD \cite{Aoki:2023qpa} and HPQCD \cite{Harrison:2023dzh} have published work on $B\to D^*\ell\nu_\ell$ form factors at nonzero recoil and HPQCD \cite{Harrison:2021tol,Harrison:2023dzh} also reported on $B_s\to D_s^*\ell\nu_\ell$. While these results are mutually compatible at high $q^2$ (see e.g.~the combined analyses in Refs.~\cite{Bordone:2024weh,Martinelli:2023fwm}), some tension is present in the slope. Directly studying the low $q^2$ region at high precision will be important for future determinations.

In this work we use RBC/UKQCD's 2+1 flavour gauge field ensembles to study exclusive semileptonic $B_{(s)}\to D_{(s)}^*\ell \nu_\ell$ decays. First we define the form factors we intend to calculate and next describe in Sec.~\ref{Sec.Lattice} the setup of our lattice calculation, before presenting first results of our analysis. For now we restrict ourselves to $B_s\to D_s^*\ell\nu_\ell$ form factors, which at this point is limited to the extraction of the form factors on one ensemble. We close by summarising and giving an outlook on our future steps.

\section{Form Factors}\label{Sec.FF}
\begin{figure}[tb]
    \centering
    \includegraphics[height=0.19\textheight]{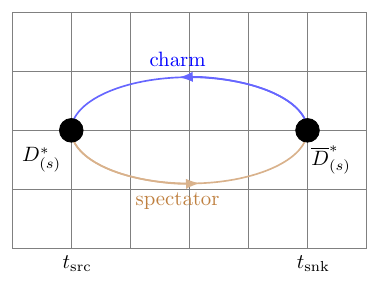}\qquad
    \includegraphics[height=0.19\textheight]{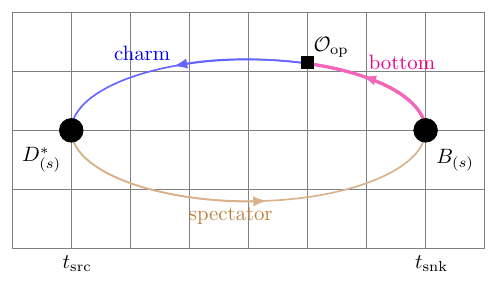}
    \caption{Sketch of the quark-line diagrams for the lattice calculation of  the 2-point functions describing the $D_{(s)}^*$ meson (left panel) and the 3-point functions for the $B_{(s)}\to D_{(s)}^*$ decay (right panel).}
    \label{fig:quarkline}
\end{figure}

Due to the narrow width of $D_{(s)}^{*}$, we approximate the $D_{(s)}^{*}$ to be QCD-stable and parametrise semileptonic $B_{(s)}\to D_{(s)}^*\ell\nu_\ell$ decays by four form factors which we define as a function of the momentum transfer $q^2$ received by the lepton pair 
 \begin{align}
  \langle D_{(s)}^*(k,\varepsilon) |\bar c \gamma^\mu b| B_{(s)}(p)\rangle &= V(q^2) \frac{2\mathfrak{i}\epsilon^{\mu\nu\rho\sigma}\varepsilon_\nu^*k_\rho p_\sigma}{M_{B_{(s)}}+M_{D^*_{(s)}}}, \nonumber\\        
        \langle D_{(s)}^*(k,\varepsilon) |\bar c \gamma^\mu\gamma_5 b| B_{(s)}(p)\rangle=& A_0(q^2)\frac{2M_{D^*_{(s)}} \varepsilon^*\cdot q}{q^2}   q^\mu \nonumber\\
    &+ A_1(q^2)(M_{B_{(s)}} + M_{D^*_{(s)}})\left[ \varepsilon^{*\mu}  - \frac{\varepsilon^*\cdot q}{q^2}   q^\mu \right] \nonumber\\
    &-A_2(q^2) \frac{\varepsilon^*\cdot q}{M_{B_{(s)}}+M_{D^*_{(s)}}} \left[ k^\mu + p^\mu - \frac{M_{B_{(s)}}^2 -M_{D^*_{(s)}}^2}{q^2}q^\mu\right],\label{eq:ff}
  \end{align}
 where $k$ and $p$ are the momentum of the $D_{(s)}^*$ and $B_{(s)}$ respectively, $\varepsilon$ is the polarisation of the $D_{(s)}^*$, $q_\mu=p_\mu-k_\mu$ is the transferred momentum, and $M_{B_{(s)}}$ and $M_{D_{(s)}^*}$ are the meson masses. We calculate these matrix elements using 2- and 3-point correlation functions implemented according to the quark-line diagrams depicted in Fig.~\ref{fig:quarkline}.

For the 2-point functions we calculate the meson-to-vacuum matrix element by inserting a colour source at Euclidean time $t_\text{src}$ and then letting the spectator strange quark propagate to $t_\text{snk}=t_\text{src}+\Delta t$ where it is annihilated with the charm quark propagating backwards in time from $t_\text{snk}$ to $t_\text{src}$. 
For the 3-point quark line diagram we start with a strange quark created at $t_\text{src}$ and propagated forward in Euclidean time to $t_\text{snk}$ where it is annihilated and turned into a sequential source for a $b$ quark which propagates backwards in time. At $t_\text{src}<t<t_\text{snk}$ we insert a weak interaction turning the $b$ into a $c$ quark which finally is annihilated at $t_\text{src}$.
To simplify the notation subsequently we set $t_\text{src}=0$ and define the 2-point functions as 
\begin{align}
    C^{\text{2pt},j}_{D_{(s)}^*}(t,k)&=\sum_{\vec{x}}\langle \mathcal{O}^j_{D_{(s)}^*}(\vec{0},0)|\mathcal{O}^{\dagger j}_{D_{(s)}^*}(\vec{x},t)\rangle e^{\mathfrak{i}\vec{k}\cdot\vec{x}}\quad\underrightarrow{t\to\infty}\quad \sum_{\lambda} \varepsilon^j(k,\lambda)\varepsilon^{j*}(k,\lambda)\frac{|\kappa_{D_{(s)}^*}|^2}{2E_{D_{(s)}^*}}e^{-E_{D_{(s)}^*}t},\\
    C^{\text{2pt}}_{B_{(s)}}(t,p)&=\sum_{\vec{x}}\langle \mathcal{O}_{B_{(s)}}(\vec{0},0)|\mathcal{O}^\dagger_{B_{(s)}}(\vec{x},t)\rangle e^{\mathfrak{i}\vec{p}\cdot\vec{x}}\quad\underrightarrow{t\to\infty}\quad \frac{|\kappa_{B_{(s)}}|^2}{2E_{B_{(s)}}}e^{-E_{B_{(s)}}t},
\end{align}
with $\mathcal{O}^\mu_{D_{(s)}^*}(x)=\overline{s}(x)\gamma^{\mu}c(x)$ and $\mathcal{O}_{B_{(s)}}(x)=\overline{s}(x)\gamma^{5}b(x)$. Hence we obtain
\begin{align}
    \kappa_{D_{(s)}^*}\varepsilon^{j*}(\vec{k},\lambda)=\langle D_{(s)}^*(\vec{k},\lambda)|\overline{c}\gamma^jq|0\rangle \quad \text{and}\quad
    \kappa_{B_{(s)}}&=\langle B_{(s)}(\vec{p},\lambda)|\overline{b}\gamma^5q|0\rangle,
\end{align}
where the polarisation vector satisfies $\sum_{\lambda}\varepsilon^{\mu}(k,\lambda)\varepsilon^{\nu}(k,\lambda)=\frac{k^\mu k^\nu}{M_{D_{(s)}^*}}-g^{\mu\nu}$ and $q=u/d,s$. The 3-point functions are defined as
\begin{align}
    &\!\!\!\!\!\!\!C_{B_{(s)}\to D_{(s)}^*}^{3pt, \Gamma,\mu}(t,t_\text{snk},k)=\sum_{\vec{x},\vec{y}}e^{\mathfrak{i}\vec{k}\cdot \vec{x}}\langle\mathcal{O}^\mu_{D_{(s)}^*}(\vec{0},0)|\mathcal{J}^{\Gamma}(\vec{y},t)|\mathcal{O}^{\dagger}_{B_{(s)}}(\vec{x},t_\text{snk})\rangle \nonumber \\
    &\underrightarrow{t,t_\text{snk}\to\infty}\quad\frac{\kappa_{D_{(s)}^*}\kappa_{B_{(s)}}}{4E_{D_{(s)}^*}M_{B_{(s)}}}\sum_{\lambda}\varepsilon^\mu(k,\lambda)\langle D_{(s)}^*(k,\lambda)|\mathcal{J}^{\Gamma}(\vec{y},t)|B_{(s)(p)}\rangle e^{-E_{B_{(s)}}t-M_{B_{(s)}}(t_\text{snk}-t)}
\end{align}
with the current $\mathcal{J}^\Gamma(\vec{x},t)=\overline{c}(\vec{x},t)\Gamma b (\vec{x},t)$ and $\Gamma=\{\gamma^\mu, \gamma^\mu \gamma_5\}$.

To extract the desired matrix elements, we form ratios of 3- over 2-point functions for which ground-state dominance in the limit of infinite time separations allows us to extract the desired matrix elements
\begin{align}
        R^{\Gamma,\mu}_{B_{(s)}\to D_{(s)}^*}(t,t_\text{snk})&=\dfrac{C_{B_{(s)}\to D_{(s)}^*}^{3pt, \Gamma,\mu}(t,t_\text{snk},k)}{\sqrt{ C_{D_{(s)}^*}^{2pt}(t,k)C_{B_{(s)}}^{2pt}(t_{snk}-t,p)}}\sqrt{\dfrac{4E_{D_{(s)}^*}M_{B_{(s)}}\sum_{j}\varepsilon_j(k)\varepsilon^{*j}(k)}{e^{-E_{D_{(s)}^*}}e^{-M_{B_{(s)}}(t_\text{snk}-t)}}}\nonumber\\
        &\qquad\xrightarrow[t_\text{snk}-t\rightarrow \infty]{t\rightarrow \infty}\varepsilon^\mu (k)\langle D_{(s)}^*(k, \varepsilon)|\overline{c}\Gamma b|B_{(s)}(p)\rangle.
    \end{align}
Going back to Eq.~\eqref{eq:ff}, we obtain the different form factors by  choosing different combinations of momenta, polarisations, and directions of currents. 
For $V$ we have to pay attention that $\varepsilon$, $k$ and $p$ are contracted with the anti-symmetric $\epsilon$ tensor and that there is no contribution for $\vec{k}=(0,0,0)$ because of the $1/k_i$ factor. 
For $A_0$ and $A_1$ it suffices to take certain combinations of directions into account to remove contributions from other form factors to the axial matrix element, whereas for $A_2$ we need to consider linear combinations of $A_1$ and $A_2$. These form factors are given by
\begin{align}
    \widetilde{V}(q^2)&=-\frac{\mathfrak{i}}{2}\frac{M_{B_{(s)}}+M_{D_{(s)}^*}}{M_{B_{(s)}}}\frac{1}{k_n}\epsilon^{0ljn}\sum_\lambda\varepsilon^j(k,\lambda)\langle D_{(s)}^*(k,\lambda)|c\gamma^l b|B_{(s)}(p)\rangle\label{eq:ffv},\\
    \widetilde{A}_0(q^2)&=\frac{1}{2}\frac{M_{D_{(s)}^*}}{E_{D_{(s)}^*}M_{B_{(s)}}}\frac{1}{k_j}q^l\sum_\lambda\varepsilon^j(k,\lambda)\langle D_{(s)}^*(k,\lambda)|c\gamma^l\gamma_5 b|B_{(s)}(p)\rangle\label{eq:ffa0},\\
     \widetilde{A}_1(q^2)&=-\frac{1}{M_{D_{(s)}^*}+M_{B_{(s)}}}\sum_\lambda\varepsilon^l(k,\lambda)\langle D_{(s)}^*(k,\lambda)|c\gamma^l\gamma_5 b|B_{(s)}(p)\rangle\label{eq:ffa1},\\
     \widetilde{A}_2(q^2)&=\frac{M_{D_{(s)}^*}\left(M_{B_{(s)}}-E_{D_{(s)}^*}\right)}{2k^lk^j\left(M_{B_{(s)}}^2E_{D_{(s)}^*}\right)}\left[\widetilde{A}_1(q^2)\left(M_{B_{(s)}}+M_{D_{(s)}^*}\right)\left(\frac{k^lk^j}{M_{D_{(s)}^*}^2}-\delta_{lj}\right)\right.\nonumber\\
     &\qquad\left.-\sum_\lambda\varepsilon^l (k,\lambda)\langle D_{(s)}^*(k,\lambda)|c\gamma^j\gamma_5 b|B_{(s)}(p)\rangle\right],\label{eq:ffa2}
\end{align}
where $1/k_i$ refers to the inverse of the $i$-component of $\vec k$.
The tilde indicates that these form factors are lattice quantities which still require renormalisation.  We plan to calculate the renormalisation factors using mostly nonperturbative renormalisation \cite{Hashimoto:1999yp, El-Khadra:2001wco} where the flavour diagonal parts are computed nonperturbatively on the lattice and the $\rho$ factor using 1-loop perturbation theory,
\begin{align}
    Z^{hl}_{J_\mu}=\rho^{hl}_{J_\mu}\sqrt{Z^{ll}Z^{hh}}.
\end{align}
The perturbative calculation is ongoing, and we intend to include at least an overall blinding factor before continuing with our analysis. The renormalised matrix elements are then given by
\begin{align}
    \langle D_{(s)}^*|\mathcal{J}_\mu|B_{(s)}\rangle=Z^{hl}_{J_\mu}\langle D_{(s)}^*|J_\mu|B_{(s)}\rangle.
\end{align}
In addition we plan to ${\cal O}(a)$-improve the operators at 1-loop for which also a perturbative coefficient needs to be calculated and additional matrix elements have already been determined as part of our lattice calculation.

\section{Lattice Setup and first results}\label{Sec.Lattice}
For our calculation we use RBC/UKQCD's 2+1 flavour gauge field ensembles with dynamical up/down and  strange quarks in the sea sector. For the light and the strange quarks we use a unitary formulation of Shamir domain-wall fermions \cite{Shamir:1993zy,Furman:1994ky}. For the $c$ quark we are using optimised heavy domain-wall fermions \cite{Cho:2015ffa,Boyle:2016imm}, whereas for the $b$ quark we use the relativistic heavy quark action \cite{Christ:2006us,Lin:2006ur}. Details of the ensembles considered are summarised in Table \ref{tab:setup} and the data analysed here were collected as part of the $B_s\to K\ell\nu_\ell$ form factor calculation presented in Ref.~\cite{Flynn:2023nhi}. Hence our data feature the same setup: light and strange quarks are simulated using point sources, while charm and bottom quarks are generated using Gaussian smeared sources. We study this quark flavour changing decay by simulating with the $B_{(s)}$ meson at rest, only injecting momenta to the final state $D_s^*$ meson. The data are taken using code written with the \texttt{Chroma} software package \cite{Edwards:2004sx}.\\
The statistical data analysis is performed using a single elimination jackknife procedure. In this first round of assessing the quality of our data only correlated ground-state fits are performed.

\begin{table}[t]
    \centering
    \begin{tabular}{l|c c c c c c c }
    \hline\hline
    & L/a & T/a & $a^{-1}$ / GeV & $am_l^{sea}$ &$am_s^{sea}$ & $M_\pi/$ MeV & srcs $\times$ $N_{conf}$\\\hline\hline
       C1 & 24 &64 &1.7848 &0.005 &0.040 &340 &1 $\times$ 1636\\
         C2 & 24& 64& 1.7848& 0.010& 0.040& 433& 1 $\times$ 1419\\
         M1& 32 &64 &2.3833 &0.004 &0.030 &302 &2 $\times$ 628\\
         M2 &32 &64 &2.3833 &0.006 &0.030 &362 &2 $\times$ 889\\
         M3 &32 &64 &2.3833 &0.008 &0.030 &411 &2 $\times$ 544\\
         F1S &48 &96 &2.785 &0.002144& 0.02144& 268& 24 $\times$ 98  \\\hline\hline     
         \end{tabular}
         \caption{RBC/UKQCD coarse (C), medium (M) and fine (F) gauge field ensembles with 2+1 flavour domain-wall fermions and Iwasaki gauge action \cite{Allton:2008pn,Aoki:2010dy,Blum:2014tka,Boyle:2017jwu, Boyle:2018knm}.} 
         \label{tab:setup}
\end{table}
    
\begin{figure}[tb]
    \centering
    \includegraphics[width=0.48\textwidth]{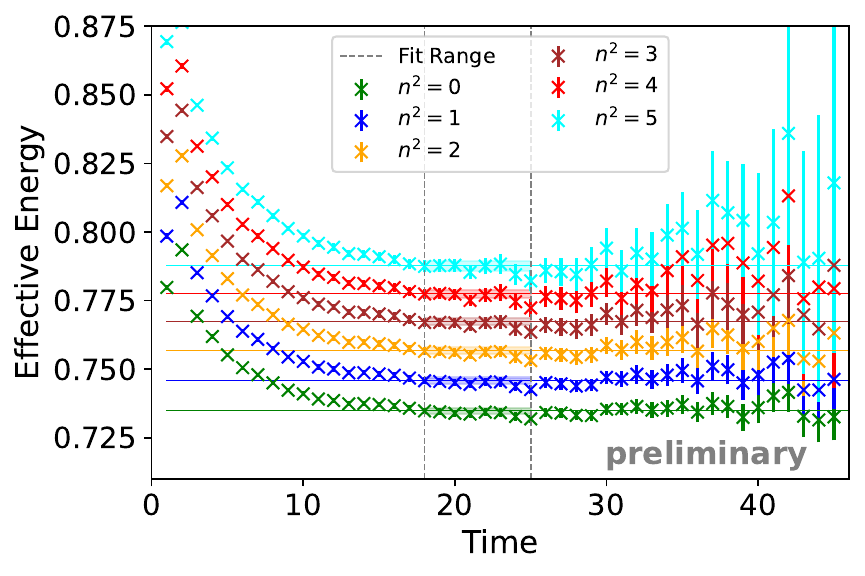}\hfill
     \includegraphics[width=0.48\textwidth]{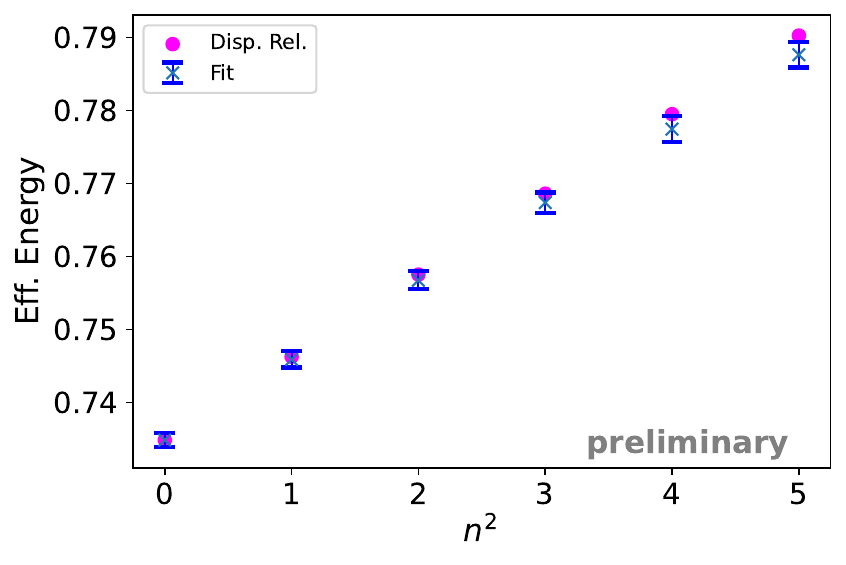}
    \caption{On the left we show the effective energies of the $D_s^*$ for different momenta in lattice units as function of the Euclidean time on the F1S ensemble. Ground state fits are indicated by the solid lines with shaded error bands between the vertical dotted lines denoting the fit range. On the right we compare these effective energies to the prediction using the lattice dispersion relation.}
    \label{fig:Dseff}
\end{figure}

We start by extracting effective $D_s^*$ meson energies from our 2-point correlators. On the one hand these enter as inputs for the extraction of the form factors defined in Eqs.~\eqref{eq:ffv}-\eqref{eq:ffa2}; on the other hand they allow us to scrutinise our setup by studying the dispersion relation. In the left plot of Fig.~\ref{fig:Dseff} we present the effective energies for a $D_s^*$ meson on our F1S ensemble with lattice momenta $p^2=(2\pi n/ L)^2$ ranging from $n^2=0$ to $5$.  Equivalent momenta in $x$, $y$, $z$ directions are averaged. Performing a ground-state-only fit, we fit time slices 18 to 25.

Using the $n^2=0$ value as input (green line in the left plot of Fig.~\ref{fig:Dseff}), we can use the lattice dispersion relation 
\begin{align}
    E=2a^{-1}\sinh^{-1}\sqrt{\sinh^2\left(\frac{am}{2}\right)+\sum_{i=1}^3\sin^2\left(\frac{ap_i}{2}\right)},
\end{align}
to check our setup. We show the comparison between the prediction from the dispersion relation to our extracted values in the right plot of Fig.~\ref{fig:Dseff} and find good agreement for all measured $n^2$ values.\\
\begin{figure}[tb]
    \centering
    \includegraphics[scale=.43]{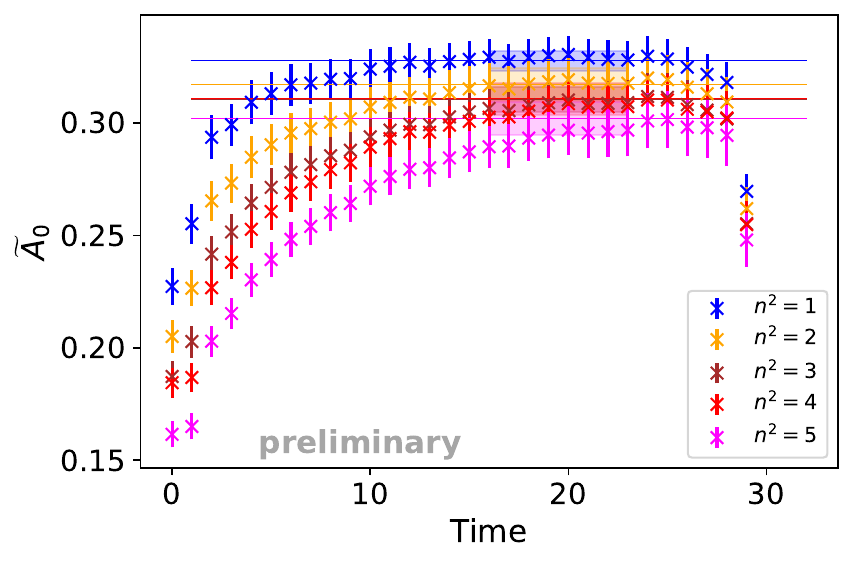}\quad
     \includegraphics[scale=.43]{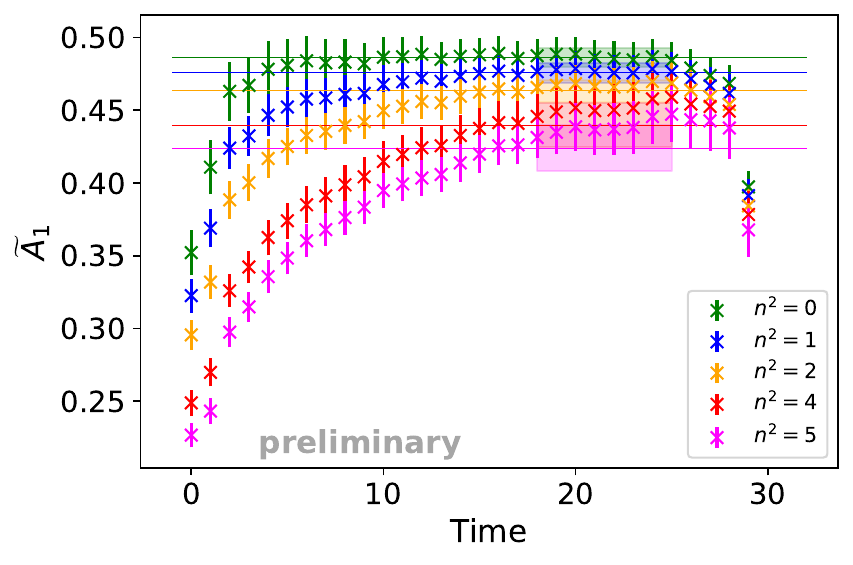}\\
     \includegraphics[scale=.43]{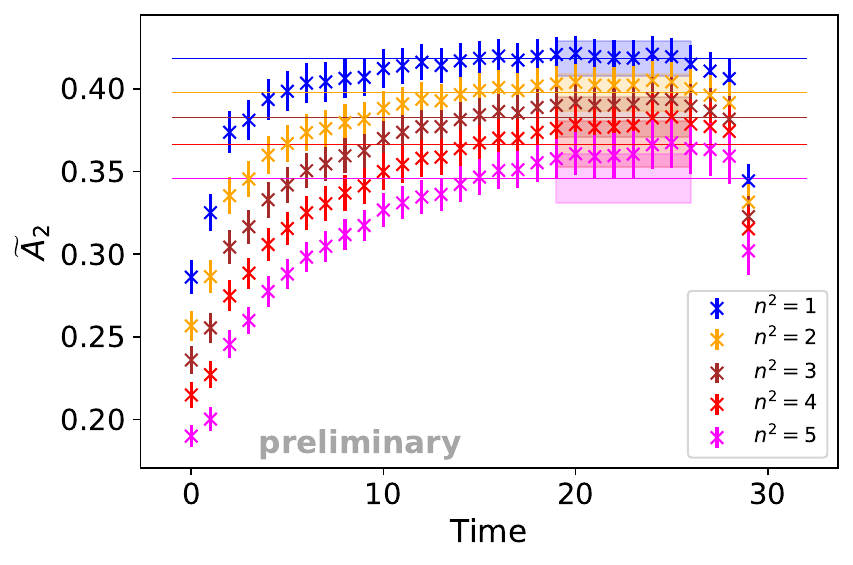}\quad
     \includegraphics[scale=.43]{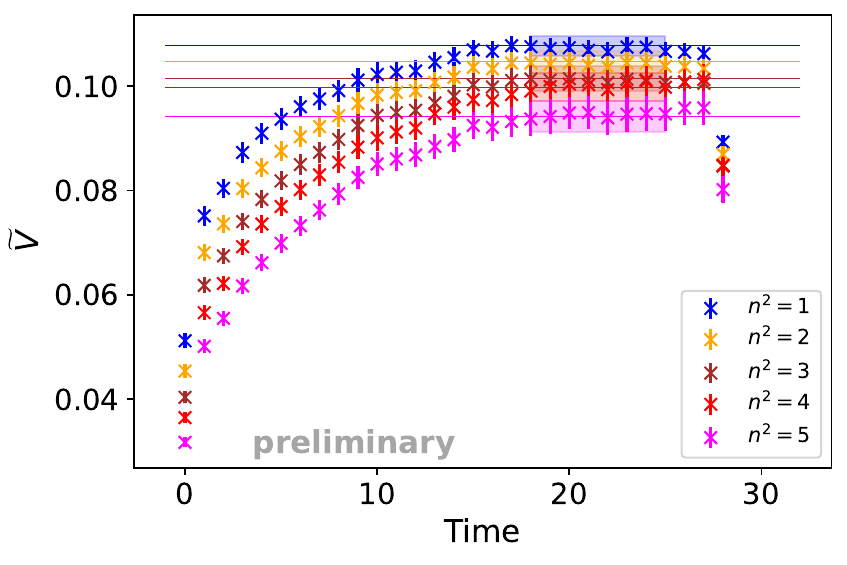}
    \caption{The four unrenormalised lattice form factors for different lattice momenta $n^2$ of the $D_s^*$ as a function of the Euclidean time on the F1S ensemble. Solid lines indicate ground state only fits with uncertainties denoted by coloured bands. \vspace{-1mm}
    }
    \label{fig:FFs}
\end{figure}
Next we proceed to extract the four form factors on F1S and show the corresponding plots in Fig.~\ref{fig:FFs}.
Again the different $n^2$ values are shown by different colours, the correlated ground state fits are given by the solid lines and the fit ranges as well as the error of the fits are indicated by the coloured band around the fit line. As discussed above, $n^2=0$ is not present for $\widetilde{V}$, $\widetilde{A}_0$ and $\widetilde{A}_2$, whereas with the $B_s$ meson at rest we have no $n^2=3$ contribution for $\widetilde{A}_1$.\\
In order to get an idea which range in $q^2$ our data covers, we convert the different lattice momenta for the F1S ensemble to physical $q^2$ values in GeV$^2$ and show our unrenormalised lattice form factors as a function of $q^2$ in Fig.~\ref{fig:qsq}.
The $q^2$ range before the chiral-continuum extrapolation is similar to but slightly smaller than the ranges by Fermilab/MILC \cite{FermilabLattice:2021cdg} and JLQCD \cite{Aoki:2023qpa}, whereas HPQCD \cite{Harrison:2023dzh} uses a different setup/analysis strategy to directly simulate the kinematically allowed $q^2$ range.  So far we have only analysed data from the finest ensemble, F1S, and are working on analysing data from medium (M) and coarse (C) ensembles. With more data included we expect our range to increase slightly and become similar to the ranges of the Fermilab/MILC and JLQCD calculations.

\begin{figure}[tb]
    \centering
    \includegraphics[height=0.21\textheight]{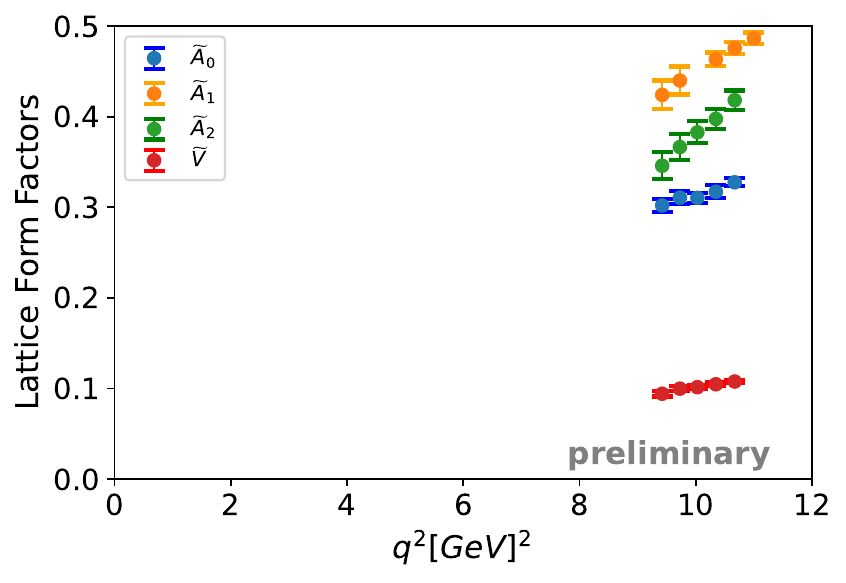}
    \caption{The fit results from Figure \ref{fig:FFs} for four lattice form factors, $\widetilde{A}_0$, $\widetilde{A}_1$, $\widetilde{A}_2$ and $\widetilde{V}$, in terms of the momentum transfer $q^2$ in physical units.}
    \label{fig:qsq}
\end{figure}
\section{Summary and outlook}
Semileptonic $B_{(s)}$ decays play a key role in determining the CKM matrix element $|V_{cb}|$. Given the persistent $2-3\sigma$ tension between the inclusive and exclusive determination, further insight is highly desired. Since, in case of exclusive decays, vector final states are experimentally preferred, we have started with exploring calculating form factors describing $B_s \to D_s^*\ell\nu$ decays. We use the narrow width approximation for the $D_s^*$ meson and focus for now on the process with a strange spectator quark before tackling $B\to D^*\ell\nu$, favoured by the $B$-factories, in the future.

Taking advantage of available data, we extract the four form factors and present their $q^2$ dependence. Keeping the $B_s$ meson at rest, we inject up to five units of momentum to the $D_s^*$ final state meson. This allows us to directly study the high $q^2$ region in a range compatible to the work by Fermilab/MILC and JLQCD.

We are analysing further available data on the coarse (C) and medium (M) ensembles listed in Tab.~\ref{tab:setup} in order to perform extra-/intrapolations to physical quark masses and the continuum in the future. This will require the $\rho$ factors needed for pursuing a mostly nonperturbative renormalisation as well as the coefficients for 1-loop ${\cal O}(a)$ operator improvement. These corresponding perturbative calculations are already in progress and are performed separately to allow for including an overall blinding factor. In parallel, work is ongoing to improve our setup and explore possibilities to directly study form factors at lower values of $q^2$ with high precision on the lattice. \vspace{-2mm}

\acknowledgments
This work was supported by the Deutsche Forschungsgemeinschaft (DFG, German Research Foundation) under Grant No.~396021762-TRR 257 ``Particle Physics Phenomenology after the Higgs Discovery''. AB ackowledges support from the House of Young Talents at the University of Siegen, Germany.
Computations used resources provided by the USQCD Collaboration, funded by the Office of Science of the
U.S.~Department of Energy and by the \href{http://www.archer.ac.uk}{ARCHER} UK
National Supercomputing Service, as well as computers at Columbia University,
Brookhaven National Laboratory, and the OMNI cluster of the University of Siegen.
This document was prepared using the resources of the USQCD Collaboration at
the Fermi National Accelerator Laboratory (Fermilab), a U.S.~Department of
Energy (DOE), Office of Science, HEP User Facility. Fermilab is managed by Fermi Research
Alliance, LLC (FRA), acting under Contract No.~DE-AC02-07CH11359.
This work used the DiRAC Extreme Scaling service at the University of Edinburgh,
operated by the Edinburgh Parallel Computing Centre on behalf of the STFC
\href{https://dirac.ac.uk}{DiRAC} HPC Facility. This equipment was funded by BEIS capital
funding via STFC capital grant ST/R00238X/1 and STFC DiRAC Operations grant
ST/R001006/1. DiRAC is part of the National e-Infrastructure. 
We used gauge field configurations generated on
the DiRAC Blue Gene~Q system at the University of Edinburgh, part of the DiRAC
Facility, funded by BIS National E-infrastructure grant ST/K000411/1 and STFC
grants ST/H008845/1, ST/K005804/1 and ST/K005790/1.

{
  \bibliography{B_meson.bib}
  \bibliographystyle{JHEP-jmf-arxiv}
}


\end{document}